# Time-modulated excitation for enhanced single molecule localization microscopy

Pierre Jouchet[1], Christian Poüs[2], Emmanuel Fort[3], Sandrine Lévêque-Fort[4*]

[1] *Université Paris-Saclay, CNRS, Institut des Sciences Moléculaires d'Orsay, 91405, Orsay, France, ORCID 0000-0002-9501-9948*

[2] *Université Paris-Saclay, INSERM UMR 1193, Châtenay-Malabry, France, ORCID 0000-0002-2502-7854*

[3] *Institut Langevin, ESPCI Paris, CNRS, PSL University, Paris, France, ORCID 0000-0003-2770-3753*

[4] *Université Paris-Saclay, CNRS, Institut des Sciences Moléculaires d'Orsay, 91405, Orsay, France, ORCID 0000-0002-9218-3363*

## Summary

Structured illumination in Single Molecule Localization Microscopy provides new information on the position of molecules and thus improves the localization precision compared to standard localization methods. Here, we used a time-shifted sinusoidal excitation pattern to modulate the fluorescence signal of the molecules whose position information is carried by the phase and recovered by synchronous demodulation. We designed two flexible fast demodulation systems located upstream of the camera, allowing us to overcome the limiting camera acquisition frequency and thus to maximize the collection of photons in the demodulation process. The temporally modulated fluorescence signal was then sampled synchronously on the same image, repeatedly during acquisition. This microscopy, called ModLoc, allows to experimentally improve the localization precision by a factor of 2.4 in one direction, compared to classical Gaussian fitting methods. A temporal study and an experimental demonstration both show that the short lifetimes of the molecules in blinking regimes impose a modulation frequency in the kilohertz range, which is beyond the reach of current cameras. A demodulation system operating at these frequencies would thus be necessary to take full advantage of this new localization approach.

## 1.Introduction

Single Molecule Localization Microscopy (SMLM) (1–3) is a fluorescence imaging method that bypasses the diffraction limit and thus allows spatial resolutions of the order of 20-50 nm, compared to several hundred nanometers for standard imaging methods. This approach of fluorescence imaging relies on the detection and localization of single molecules (i.e. the only ones to emit in a focal volume) that emit randomly over time. The molecules present in the sample are then in a regime called blinking regime which can be obtained from different modalities such as the (DNA)-PAINT (for Points Accumulation for Imaging in Nanoscale Topography)(4–7), the (d)-STORM (for (direct)-Stochastic Optical Reconstruction Microscopy) (3,8)or PALM (Photo-Activated Localization Microscopy) (1,2). The localization of a fluorescent probe is characterized by its accuracy and by its precision of localization. The precision depends essentially on the model used to localize the fluorescent probe. Under standard SMLM conditions, the use of large numerical aperture objective (9) has shown that a single molecule image (PSF) adjustment could be performed using a simple 2-D Gaussian function. Based on this model, analytical formulas have been proposed to estimate the localization precision associated with the maximum likelihood estimator (10,11) which shows its strong dependency on the spatial properties of the PSF. Other alternative localization methods based on the use of the spatial properties of illumination have also been developed to perform single molecule tracking (12,13) of single polystyrene bead in a travelling interference pattern, thanks to a photodiode and a lock-in detection. This concept has been revisited recently for SMLM imaging in order to improve lateral precision (14–16) or axial precision (17). These new approaches rely on the variation of the fluorescence signal of a fluorescent probe when it is illuminated with a moving structured illumination. The localization is obtained from the phase measurement given by the lock-in detection of the modulated fluorescence signal. This microscopy, called ModLoc in the following, is a wide field approach. The time-modulated excitation is identical for all emitters in the field of view (FOV), contrary to MINFLUX (18) for which the sequence of excitations is optimized for a position in the FOV.

*Author for correspondence (sandrine.leveque-fort@université-paris-saclay.fr)

Retrieving the position of a fluorescent probe from a moving illumination pattern can be performed with a sinusoidal illumination pattern commonly used in Structured Illumination Microscopy (SIM) (19–21). This pattern is shifted in time at constant speed to modulate the fluorescence signal of single emitter. The range of modulation frequencies is determined by the demodulation strategy which allows the determination of the position of the fluorescent probe within the pattern. A sequential images acquisition strategy is limited by the maximum acquisition rate of the camera. Alternatively, a dedicated demodulation system placed in front of the camera enables fast demodulation frequencies that can be optimized with the blinking kinetics of the emitters. In this context, here we propose a new implementation in the lateral direction of our fast demodulation ModLoc strategy previously used along the axial direction. We show that it allows an experimental improvement of the transverse localization precision by a factor of 2.4 compared to classical Gaussian fitting methods. We also present an analysis of single molecule emission time which motivates the development of a frequency flexible lock-in detection. This allows ModLoc to take into account the random temporal behaviour of fluorescent probes in order to be compatible with all SMLM strategies.

## 2-Principle

Let us consider a moving sinusoidal excitation pattern in the transverse direction $x$ with a shifting angular frequency $\Omega$ and a pitch $\Lambda$. The excitation pattern is given by

$$I(x,t) = I_0 \left( 1 + m \cos\left(\frac{2\pi}{\Lambda}x + \Omega t\right)\right) \quad (1)$$

with $I_0$ being the average light intensity over the whole field and $m$ is the contrast of the illumination pattern. In the linear regime, i.e. out of saturation, the fluorescence signal of a molecule is proportional to the light intensity it receives. The photon flux $N(x,t)$ detected from a fluorescent emitter located illuminated by the excitation field of the equation ( 1 ) placed at position $x$ is given by

$$N(x,t) = N_0 ( 1 + m \cos(\Phi(x) + \Omega t)) \quad (2)$$

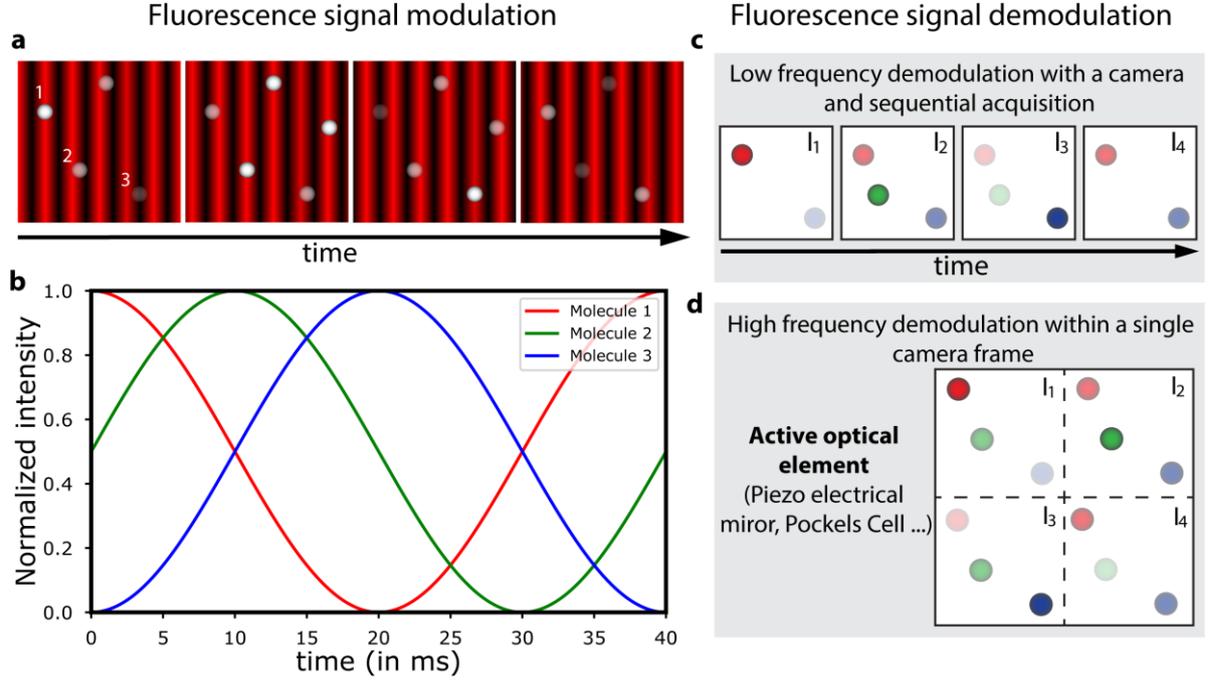

**Figure 1** : **Principle of the single molecule localization by temporal signal modulation a**. Temporal modulation of different single probes in a structured illumination. **b**. Normalized intensity of the different probes shown in **a** as a function of time. **c.** Scheme of a frame-by-frame demodulation of 3 single molecules using the camera frame rate as the demodulation frequency. One of these molecules stops emitting before the end of the modulation and cannot be used for the reconstruction. **d.** Scheme of the demodulated signal of 3 single molecules in a single camera frame. All the molecules are used for the super resolved image reconstruction.

With $\Phi(x) = \frac{2\pi}{\Lambda}x$ and $N_0$ being average detected photons flux associated to an illumination intensity $I_0$. The time-modulated signal then carries information about the position of the molecule within its phase $\Phi(x)$. Figure 1a schematized the temporal variation of intensity of different fluorescent molecules located in a sinusoidal excitation field. The variation of the different signals is then presented in Figure 1b. It can be seen that the molecules located at different positions in the field have different phases. These phases can then be obtained by demodulation with only 3 or 4 sampling points of the signal. The Cramer Rao lower bounds (CRLB), which gives access to the theoretical localization precision accessible to this method, can then be calculated from this model. In the case of a 4-point demodulation, without background noise signal and for a maximum excitation contrast, the theoretical localization precision takes the simple form:

$$\Delta_x = \frac{\sigma}{\sqrt{N(1 + \frac{4\pi^2 \sigma^2}{\Lambda^2})}} \quad (3)$$

Note that the spatial properties of the excitation pattern are reported in the location precision. Thus, the localization precision worsens upon increasing the spatial period. The CRLB for different values of the pattern parameters can be compute numerically and the details of calculation are presented in the Supplementary Note 1.

The temporal characteristics of the fluorescence signal imposes specific constraints on the demodulation strategies. Different studies (14,16) proposed a frame-by-frame demodulation method. The latter, presented in

Figure 1c, is based on the acquisition of an image corresponding to each position of the excitation pattern. This demodulation strategy is limited by the acquisition frequency of the camera used, which can vary depending on the size of the observed field and the type of detector used (sCMOS or EMCCD). However, this sequential approach restricts the application to SMLM imaging. In the blinking regime, molecules start and stop emitting at random times during the acquisition. Therefore, the frames corresponding to the appearance and extinction of the fluorescence signal cannot be taken into account in the demodulation signal in order not to bias the result. Thus, the frame-by-frame demodulation induces a loss of photons degrading the accuracy of localization that could be achieved initially. In addition, a part of the molecules do not emit for long enough periods to be localized, as shown in Figure 1c. These molecules cannot be taken into account in the reconstruction of the super-resolved image. To minimize the proportion of molecules that must be discarded, the average ON-time must be much longer than the acquisition time needed for localization. This significantly hinder this sequential strategy which becomes time consuming.

A solution to perform demodulation at high frequencies independent of the limited camera acquisition rate consist in introduce a fast frequency demodulation system before the camera and demodulate the fluorescence signal within the same frame. As shown in figure 1e, the signal from the same probe is divided into 4 PSFs having 4 different intensities according to the 4 pattern positions. This demodulation system can be obtained by using mirrors on a piezo electrical mount or electro-optical elements. The different experimental configurations and the performances of the methods will be discussed below. Figure 2 shows the interest of the interest of modulating the fluorescence intensity to localize single molecules. The theoretical localization precision was calculated for different number of emitted photons, and pattern contrasts and different background noise. The variation of the CRLB as a function of the number of photons emitted by a single molecule and direct comparison with the gaussian fitting are shown in figure 2a. It can be noted that the theoretical localization precision rapidly drops below 2 nm for the standard photon number used in SMLM (more than 1000 detected photons per molecule). The enhancement is nearly independent in respect to the number of detected photons and only slightly sensitive to the presence of noise. The localization precision can be enhanced by a factor up to 3 with ModLoc approach. The curves in Figures 2b and 2c show the influence on the localization precision of the illumination pitch and contrast respectively, the CRLB calculation of the standard Gaussian method is represented by a black curve. We can see that the spatial pitch plays a central role in the localization precision of the method. When the spatial pitch decreases, the localization precision of ModLoc tends towards the localization accuracy of the Gaussian fitting. As the spatial period increases, the position information encoded in the phase of the fluorescence signal is no longer precise enough and the position information encoded in the PSF fitting becomes the main source of information as expected from the limit case of a uniform illumination. We note for instance that for a pitch of approximately 600 nm easily accessible with high numerical aperture objectives, the localization precision is enhanced by a factor 2. The contrast in the modulation of the illumination pattern is central. In the absence of background noise, a contrast of 0.5 divides by approximately two the theoretical localization precision of ModLoc. As expected, this precision is given by the Gaussian fitting as the contrast decreases to zero and the illumination becomes homogeneous. These performances can be achieved by a demodulation obtained with sequential acquisition. However, the temporal aspect of the fluorescence emission in single molecule localization microscopy must be considered in order to make the most of this detection strategy. A minimum of 3 samples of the modulated intensity are necessary to get the position of the single probe. As the emitter can emit or go in a dark state at any time, the molecule must be present on a minimum of 5 frames to be considered. This condition allows us to be sure that the molecule is continuously active over 3 frames. Figure 2d shows a simulation for the number of molecules fulfilling this condition as a function of the integration time of the camera (for a readout time of 1.25 ms). Simulations reveal that the frame-by-frame demodulation strategy leads to a loss of more than 30 percent of the detected molecules for an integration time of 5 ms, which decreases the quality of the final super resolved image, and as previously observed for sequential spectral unmixing.

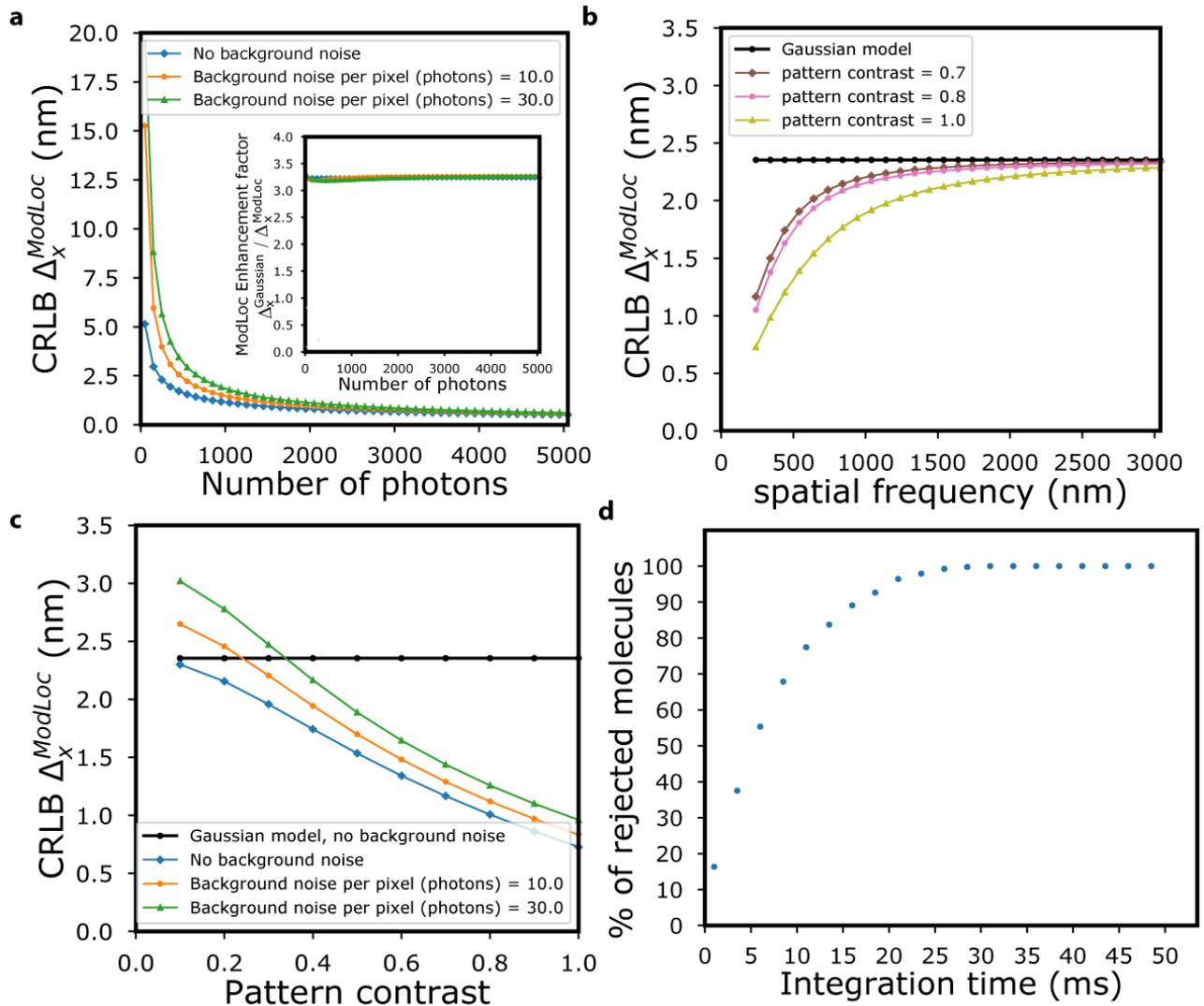

**Figure 2 : Theoretical performances of localization by modulation/demodulation of the fluorescence signal and comparison with the classical Gaussian fitting method. a.** Cramer Rao lower bounds (CRLB) as a function of the number of detected photons for different background noises. The pattern contrast is set to 1 and the spatial pitch is set to 240 nm. The comparison of the CRLB for the modulation/demodulation method and the Gaussian fitting as the function of the number of photons for different background noises is also represented in the graph. **b.** CRLB as a function of the spatial pitch for different values of the pattern contrast. Calculation are performed for 2500 detected photons and no background noise. **c.** CRLB as a function of the of the pattern contrast for different background noises. Spatial pitch is 240 nm and the number of detected photons is 2500. **d.** Percentage of molecule rejected for the modulation analysis as a function of the integration time (for a readout time of 1.25 ms). The molecules are rejected if they are not present at least on 5 successive frames.

# 3- Experimental set-up

Experimentally, the illumination pattern was obtained by two-wave interference. Figure 3 presents the experimental setup to produce a shifting illumination pattern as well as its operating principle. A laser source (Genesis MX 607/639 STM, Coherent) is divided into two beams with a 50-50 beamsplitter cube. Each beam then passes through an electro-optical phase modulator (EO-PM-NR-C1, Thorlabs) used to tune the optical phase between the two beams. The beams are then set parallel using a knife edge mirror and are focused in the back focal plane of the objective. The two beams are collimated on the sample and interfere in the object plane of the objective. The spatial pitch of the interference pattern can be adjusted by changing the position of the two beams in the back focal plane by moving the knife edge mirror mounted on a translation stage, as shown in Figure 3a. The contrast of the fringe pattern can be maximized by modifying the orientation of two half-wave plates combined with a polarizer to adjust the intensity of each beam in the sample.

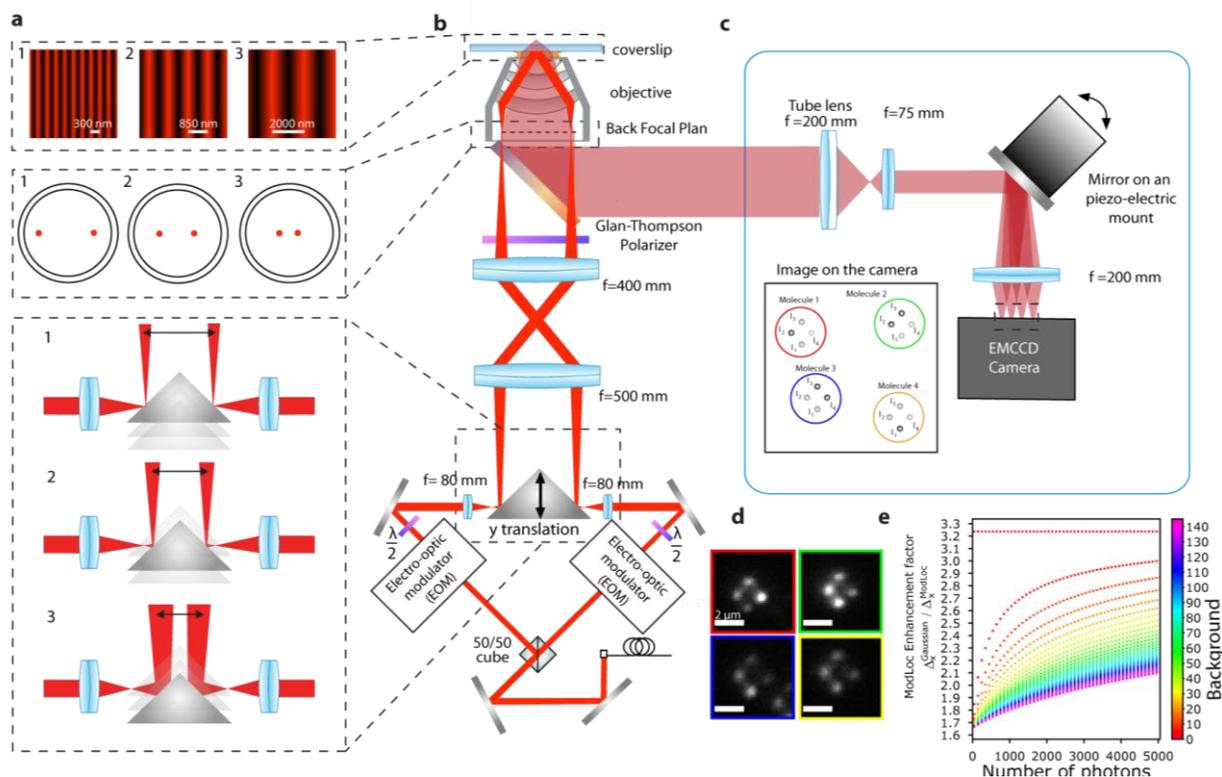

**Figure 3 : Experimental set-up for the modulation/demodulation of the single molecule fluorescence signal. a.** The fringe pattern can be easily adjusted thanks to a knife edge mirror placed on a x/y translation stage, impacting the position of the focus beam in the back focal plane and the pattern frequency on the sample plane. **b.** The laser source is fibered to increase the stability, it is divided into two beam paths thanks to a 50-50 cube beamsplitter. In each path, an electro-optical modulator allows to introduce a time variation of the positions of the fringe pattern. Two $\lambda/2$ plates and a glan Thompson polarizer allows to optimise the contrast of the pattern. **c.** The detection system is composed of a mirror placed on a piezo-electric x/y tip-tilt, which introduces a shift of the image position in synchronization with the position of the excitation pattern. Four shifted replica are thus registered on a single camera frame to sample the modulation emission. **d.** Example of PSF acquired by the camera on 40 nm fluorescent beads. **e.** Enhancement factor provided by ModLoc calculated by comparing the precision obtained with the classical Gaussian MLE and ModLoc for a demodulation with a piezo electric mirror for various background noise values. The background noise for ModLoc is higher in this configuration thus the enhancement factor is reduced when background increases.

Two demodulation modules with different assets are used to perform the lock-in detection.

A first demodulation module is shown in Figure 3c. This demodulation system is based on the use of a mirror placed on a piezoelectric support (S 330.2 SL, Physik Instrument) allowing an active deflection of the incident fluorescent beam. Each position of the illumination pattern is associated to a different position of the mirror thus spatially separating the temporal information on the same camera frame (camera EMCCD iXon 3 ANDOR). The mirror is set at a fixed position for a given illumination pattern and quickly changes position as the pattern shifts to another position. This operation is repeated 3 times for an acquisition time of 50 ms and a demodulation frequency of 60 Hz. The effective PSF of a fluorescent molecule then appears as 4 PSFs forming a diamond shape and containing the 4 samples of intensity used for demodulation. An example of the experimental PSFs of 40 nm diameter red nanobeads located at the coverslip is shown in Figure 3c. We can then observe the different intensity distributions of the beads present in the samples at different phases. We note that in this configuration, the background noise is not divided by the number of intensity samples (4 samples for this setup). This results in a diminution of the performances of the approaches. A new comparison with the gaussian fitting is shown in figure 3d for different background values.

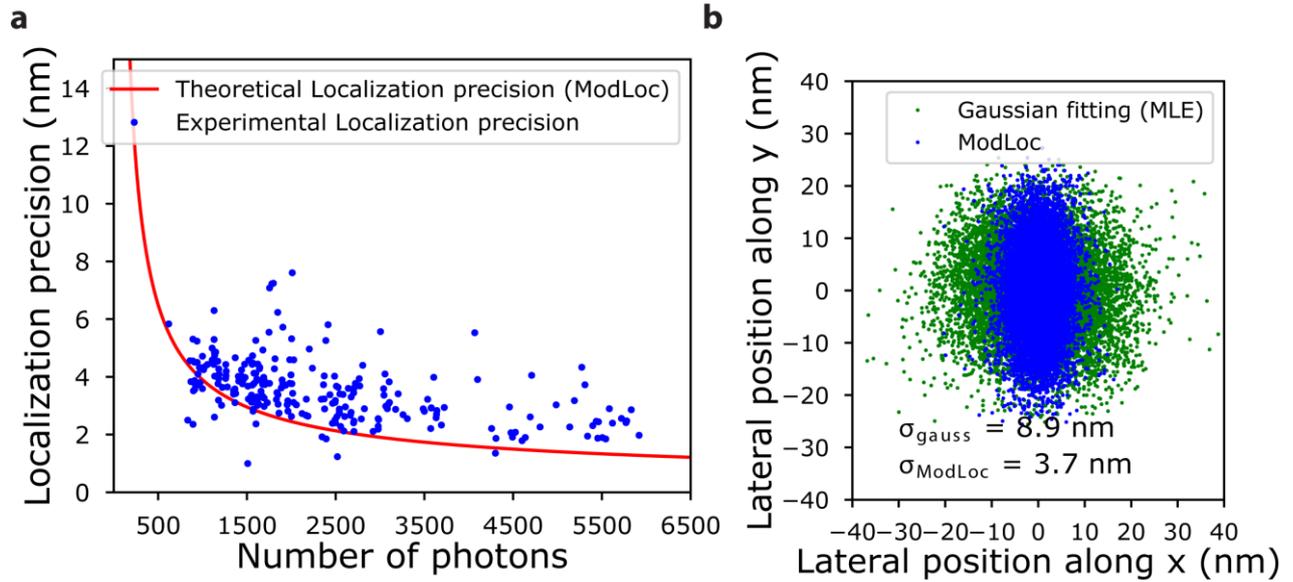

**Figure 4: Experimental localization precision of ModLoc and comparison with the classical Gaussian fitting method. a.** Localization precision as a function of the number of photons obtained from isolated red fluorescent nanobeads. Spatial pitch is 240 nm and contrast 0.9. **b.** Localization of different red fluorescent nanobeads obtained with ModLoc and with the classical Gaussian Fitting. The average position of each bead is set to zero.

## 4-Results

In order to estimate the localization precision achieved by this ModLoc implementation, we have carried out several successive acquisitions of the same sample of beads located at the coverslip (22–24). The excitation power was adjusted to obtain a fluorescence emission intensity in the range of that commonly observed with molecules in SMLM imaging. Figure 4a shows the variation of the localization precision as a function of the photon number (blue dots). The localization precision obtained from the standard deviation of the positions on each nanobead. The red curve represents the theoretical localization precision obtained from the CRLB. The spatial pitch of the illumination pattern is 240 nm and the contrast is around 0.9 (cf Supplementary figure 2). We note that the experimental results follow the theoretical expectations, showing the performances of this ModLoc implementation. Figure 4b shows the localizations obtained by ModLoc (blue dots) and by a Gaussian fitting of the PSF (green dots). The localization measurements are performed on several 40 nm-diameter red nanobeads and repeated several times on the same bead. The average position of each nanobead is set to zero by subtracting the localization mean value calculated from the localization measurements of this nanobead. The localizations obtained by Gaussian fitting have been performed with the MLE method using the GPUFIT (25) code for an isotropic 2D Gaussian model. The localizations obtained by ModLoc correspond to the x direction. The superposition of the two types of localizations allows a direct visualization of the enhancement brought by the ModLoc method along the x axis. The calculated localization precision associated with the Gaussian method is 8.9 nm while that one associated with the ModLoc method is 3.7 nm, corresponding to an enhancement by a factor 2.4. It can be noted, however, that the weak aberrations generated by the mirror deflexion at detection can degrade the Gaussian localization accuracy, and thus increase experimentally the improvement of the ModLoc method compared to the Gaussian fit.

The localization precision can be measured using 40 nm-diameter red fluorescent nanobeads because it is possible to repeat the localization on several successive acquisitions. However in SMLM, the fluorescent molecules are in a blinking regime which prevents this type of measurement. Since ModLoc introduces a temporal information in the fluorescence signal of single probes with a modulation frequency $f_{mod}$, a temporal study of the fluorescence of the probes in SMLM condition appears necessary to estimate the impact of this frequency. In the blinking regime, the molecules begin to emit at a random time and for a random ON-time $\tau_{ON}$. This duration depends on the imaging modality used and can generally be modified by adjusting the excitation power. DNA-PAINT imaging methods, for example, provide long $\tau_{ON}$ because samples are generally excited at low powers. In this case, the single molecule regime is achieved by changing the emitter concentration in the sample. In (d)-STORM imaging, the fluorescent probes are already present in the sample and the single molecule regime is achieved by adjusting the excitation power so that the molecule passes into a dark non-fluorescent state. $\tau_{ON}$ can therefore be modified linearly with the power (8). For dense samples, a high excitation power is necessary to reach the single molecule regime, thus imposing the average $\tau_{ON}$ of the molecules. Figure 5a represents the average $\tau_{ON}$ of 2 different fluorescent probes in (d)-STORM imaging, Alexa 647 (AF647) and CF660C. These characteristic times are below 20 ms and impose us acquisition times $\tau_{int}$ around 30 to 50 ms in order to collect a maximum of photons. We have to take into account these aspects in order to establish the adequate modulation/demodulation frequency.

As we control the integration time ($\tau_{int}$) of the camera and the modulation/demodulation frequency of our setup, we define $N_{cycles}$ as the number of period modulated during an acquisition. $N_{cycles}$ is thus given by

$$N_{cycles} = \tau_{int} \times f_{mod} \qquad (4)$$

We can estimate the influence of $N_{cycles}$ using simulations. The integration time being directly adapted to the ON time of the fluorescent molecules, the modulation/demodulation frequency must be adapted to keep the number of demodulation cycles constant. We simulated a fluorescent probe emitting at any time for a duration $\tau_{ON}$ whose probability density is given by the adjustment of the AF647 histogram presented in Figure 5a. The localization precision as well as the accuracy of measurement are obtained by computing the standard deviation and the mean of the position distribution for 50 simulated blinking events. This operation is then repeated for different modulation cycles. Figure 5b shows the simulation results for 3 values of $N_{cycles}$, with an integration time of 50 ms and 2500 emitted photons. The details of the simulations are given in Supplementary note 2 and Supplementary Figure 1. Low modulation frequencies result in measurement biases and poor localization precision. The evolution of the localization precision (standard deviation) as well as the accuracy (difference between the actual position and the average position computed) as a function of $N_{cycles}$ is shown in Figure 5c. For $N_{cycles} > 60$, the results tend towards the theoretical predictions. For an integration time of 50 ms, 40 modulation cycles correspond to a modulation frequency $f_{mod}$= 1200 Hz. These simulations allow to define the minimum number of demodulation cycles to obtain the best performances for ModLoc. The demodulation frequency can be estimated as a function of the integration time of the camera, itself adapted to the ON time of the molecules. The use of piezo electric devices such as in the implementation shown in Fig. 3 is possible only if set at their resonance frequency. Out of resonance, these mechanical actuators have a very slow rise time hindering large deflections. Each piezo actuator has a specific resonance frequency which can be of a few kHz. The use of the resonant mode imposes the modulation/demodulation frequency and does not allow flexibility. The optimization of the modulation frequency to the emission properties of the fluorescent probes is however most desirable. In addition, the excitation signal must be switched off when the demodulation system spatially switches from a sub region of the camera to another in order to avoid a spurious signal. We therefore propose another demodulation method based on polarization modulation using a Pockels cell enabling tunable high operating frequencies (17,26,27). This implementation, shown in Figure 6a, needs that the fluorescent probes have sufficient degrees of freedom so that their emission is not polarized. This condition is fulfilled for most of the fluorophores (28). The fluorescence signal is split in two by a 50/50 polarizing cube to obtain two equally intense cross-polarized paths. Their polarization is then modulated using a Pockels cell (CF1043, Fast Pulse). A quarter-wave plate is placed in one of the two paths. A 50/50 polarizing cube eventually creates 4 detection paths with quadrature time modulation on the camera (EMCCD iXon 3 ANDOR). The observed field on the camera is composed of 4 sub-regions corresponding to the 4 channels each associated to a position of the illumination pattern. In order to study the influence of the number of cycles on the localization precision and validate the results obtained by simulations, we used 40 nm spaced nanorulers observed at several modulation frequencies. A schematic of the sample geometry is shown in Figure 6a. The sample was observed in DNA-PAINT condition which enables long ON-times allowing us to operate with long acquisition times. The number of modulation cycles can thus be tuned over a large range of values. Different reconstructed images for different numbers of modulation cycles are represented in Figure 6b. The 3 binding sites of the nanoruler are indistinguishable at 10 modulation cycles. They appear much more clearly as the number of modulation cycles is increased. Figure 6c shows the localization standard deviation $\sigma^{NanoR}$ for single binding site as a function of the number of modulation cycles. $\sigma^{NanoR}$ is related to the localization precision of the system. We note that the evolution is identical to that obtained with the simulations and a plateau is observed from 60 cycles of modulations. This demodulation implementation thus gives access to high modulation frequencies needed in dSTORM imaging conditions. However, the polarization cross-talk between the detection channels induces a loss of contrast in the demodulation process which is reduced to 0.5. When taken into account in the theoretical value of the localization precision (without background noise and for a pattern contrast corresponding to 1) one obtains

$$\Delta_x = \frac{\sigma}{\sqrt{N(1 + \frac{\sigma^2 \pi^2}{2\Lambda^2})}} \qquad (5)$$

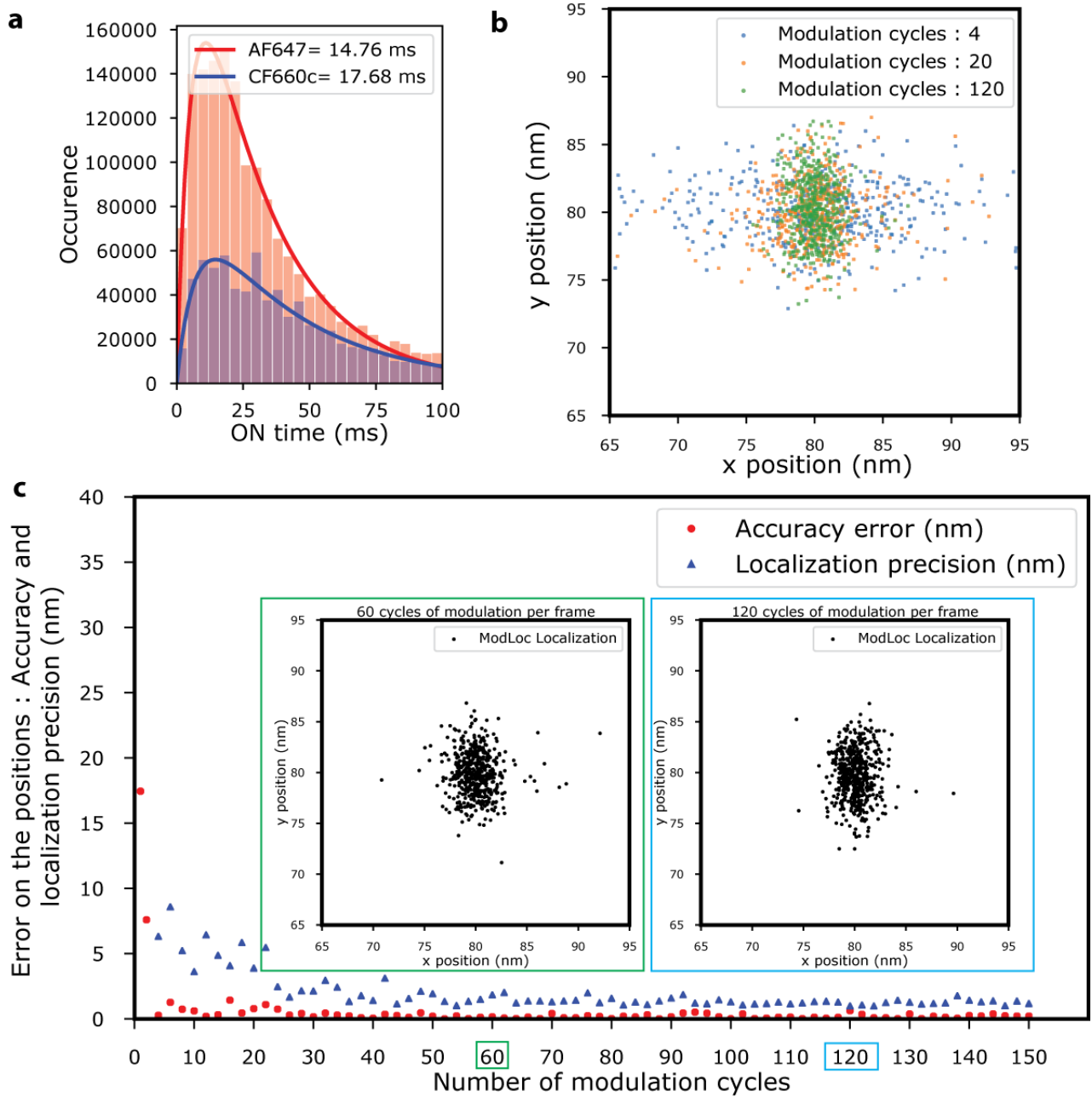

**Figure 5: Experimental data and simulations on temporal behaviour of single molecules in dSTORM imaging. a.** Experimental measurements of ON-time distribution of AF647 and CF660C acquired in classical dSTORM conditions. **b.** Localisation based on the modulated intensity of a simulated single emitter for different number of modulation cycles. ON-time of each blinking event is given by the AF647 density function (in **a.**). Each emitter is active 50 times. The number of photons emitted is 2500, the integration time is 50 ms**. c.** Localization precision and accuracy as a function of the number of modulation cycles extracted from simulations. Simulation parameters are the same as in b.

This results in a loss of performances in terms of localization precision compared with the piezo electric mirror configuration. The comparison with the gaussian fitting is shown in figure 6d .

# 5-Discussion

Our study shows that the modulation of the fluorescence signal of a single molecule allows to experimentally improve the localization precision by a factor of 2.4 compared to classical Gaussian fitting methods. ModLoc can be easily implemented using sinusoidal excitation obtained by two-wave interferences. The use of standard electro-optical components such as phase modulators allows one to accurately control the position of the illumination pattern and set the modulation frequency up to very high values. Unlike PSF adjustment methods that only take into account the spatial characteristics of the fluorescence signal, ModLoc introduces a new temporal component in the fluorescence signal. This makes ModLoc less sensitive to optical aberrations which is a real asset for in depth measurements (29,30).

The emission characteristics of the fluorophores must be taken into account in order to optimize this new localization strategy. The two experimental setups proposed in this work allow the demodulation of the fluorescence signal by two different strategies, with different advantages and drawbacks. The development of each device can be considered according to the imaging modality used to obtain the single molecule emission. The demodulation by deflection of the fluorescence signal realized by a mirror on a piezo electric support allows to obtain better performances in terms of localization precision, but leads to an increase of the effective size of the PSF which however preserves a large field of view. If the demodulation frequency is limited for this approach, however DNA-PAINT imaging methods seem to be perfectly adapted to this strategy. Indeed, increasing the effective size of a PSF requires a decrease in the density of active molecules in the sample, easily obtained in DNA-PAINT imaging. Furthermore, the ON times of the active molecules being generally much longer than in dSTORM imaging, the demodulation frequencies accessible to this approach can be adapted to this imaging modality. These time characteristics depend essentially on the imaging modality used. For example, the emission durations of fluorescent molecules in DNA-PAINT imaging enable modulation and demodulation of the fluorescence signal at low frequencies compatible with mechanical deflectors even out of resonance. On the contrary, the emission durations of fluorescent probes in dSTORM imaging are much shorter or with the recent fast DNA-PAINT development which in particular reduces drift issues (31), fast and optimized demodulation frequencies are required. The use of resonant mode deflector elements can then be used to increase the demodulation frequency but this approach limits the flexibility of the method. In addition, it requires to turn off the excitation when switching from one detection path to another. The use of a demodulation strategy based on polarization sorting represents an alternative. While limited to unpolarized samples and possibly endowed with a reduced localization precision, this strategy offers significant assets with no moving parts and a high flexibility in terms of operating frequencies, especially since most samples can be labeled with dyes that lead to unpolarized emitted fluorescence.

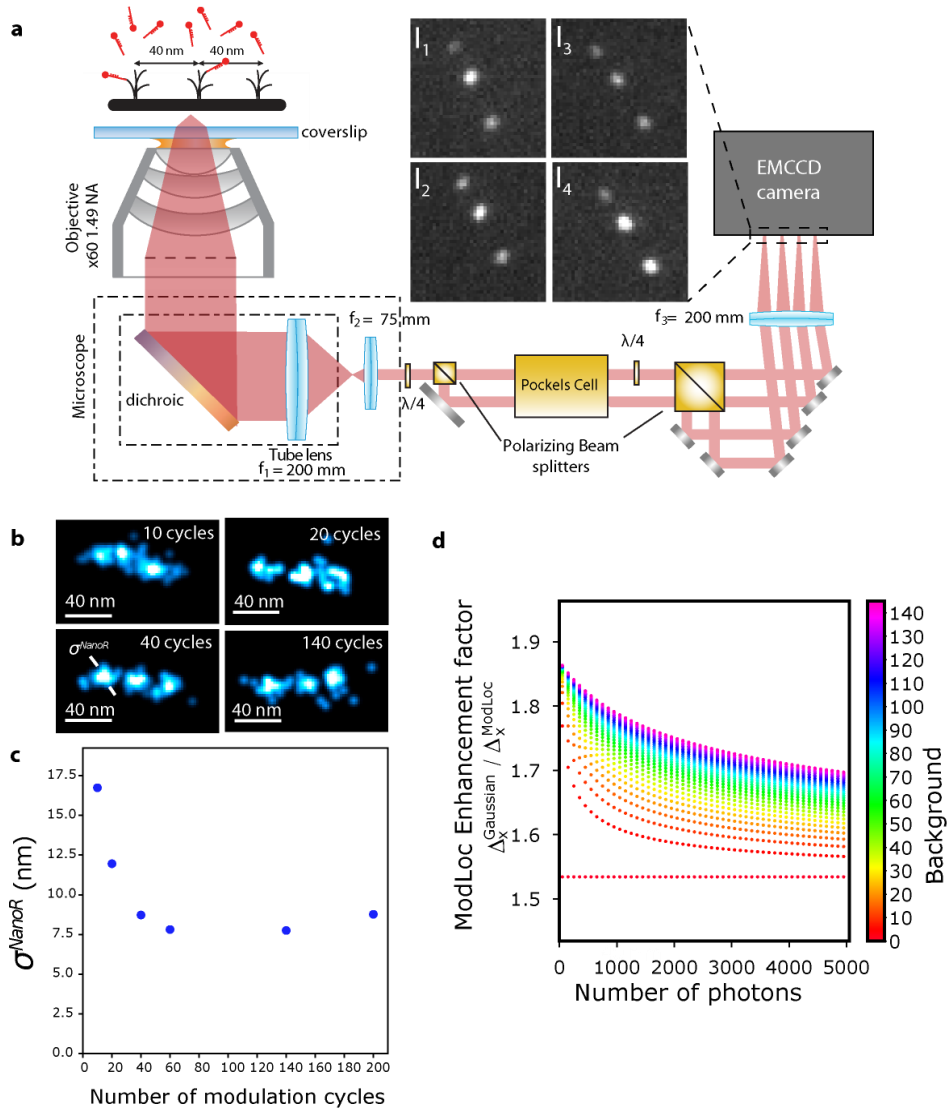

**Figure 6 : Fast demodulation setup and experimental influence of the number of modulation cycles. a.** Fast demodulation system based on the use of a Pockels cell located between 2 polarizing beam splitters. The demodulation system creates 4 images ($I_1$, $I_2$, $I_3$, $I_4$) of the same sample at different time on the camera as shown by the 4 images of 3 modulated PSF. **b.** Reconstructed images of nanorulers schematized in **a** and obtained by ModLoc with the fast demodulation system for different modulation cycles. **c.** Thickness of reconstructed nanoruler sites as the function of the modulation cycles. d. Enhancement factor provided by ModLoc calculated by comparing the precision obtained with the classical Gaussian MLE and ModLoc with the demodulation by an active Pockels cell for different background values. In this configuration, the effective contrast is reduced to 0.5 thus the enhancement factor decreases.

# 6-Conclusion

ModLoc possess several assets as an alternative localization strategy over Gaussian fitting, in particular it has an increased localization precision of more than 2 folds. However, the limited and random on-time of each emitter represents a challenging constraint which must determine the demodulation strategy to be implemented. In particular, several modulation cycles must be performed during an average on time to reduce the localization error and decrease the proportion of emitters to be discarded. A fast demodulation module based on a time sorting placed in from of the camera should be preferred to a slow sequential strategy. We have compared and discussed the assets of two complementary approach, one based on a moving mirror and one on a polarization sorting.

As the introduction of a temporal parameter in the localization strategy is an emerging field, new alternative configurations will most certainly allow the concept to move forward, while preserving the maximum number of detected molecules and even increase localization performances. This time modulation strategy is not limited to transverse localization measurements and has been proposed for the axial localization. In addition, this concept is not limited to encode the localization information but can be generalized to any parameter at the single molecular level such as orientation or lifetime (27). While, at first sight, multiplexing this lock-in strategy seems relatively straightforward and appears as a real asset, the strong constraints on the time characteristics of the emitters as well as the simultaneous measurements at various frequencies on a single emitter appears quite challenging. Smart implementations are still needed to perform efficient multiplexing which for instance would optimize the information extracted from each detected photon or would reach very high modulation frequencies for lifetime measurements.

# Additional Information

**Information on the following should be included wherever relevant.**

**Acknowledgments**
We thank Abigail Illand, Clément Cabriel, Nicolas Bourg and Guillaume Dupuis for discussion. We acknowledge the support of NVIDIA Corporation with the donation of the Titan Xp GPU used for this research, and CPBM/ISMO-UMR8214 for access to wet lab/cell culture facility. We thank Abbelight for the free use of NEO software and dSTORM buffers.

**Funding Statement**
P.J. acknowledges a master funding from GDR ImaBio, and PhD funding from IDEX Paris Saclay (ANR-11-IDEX-0003-02). This work was supported by the AXA research fund, the ANR (LABEX WIFI, ANR-10-LABX-24), ANR MSM-Modulated super-resolution microscopy (ANR-17-CE09-0040), the valorization program of the IDEX Paris Saclay and of Labex PALM (ANR-10-LABX-0039-PALM).

**Data Accessibility**

Localization data represented in figure 4 and 6 are provided as supplementary materials.

**Competing Interests**

The CNRS has deposited a patent FR3054321-A1 on the 25 July 2016 to protect this work, currently under international extension. S.L.F and E.F. are co-inventors.

**Authors'Contributions**
P.J, C.P., E.F. and S.L.F. conceived the project. P.J. designed the optical setup, performed the acquisitions, data analysis, Simulations and CRLB calculations. All authors have contributed to the manuscript.